\documentclass[aoas,MSNbibl,nameyear,seceqn,rotating,dvips]{arximspdf}
\usepackage{dcolumn}
\usepackage{graphicx}

\doi{10.1214/11-AOAS515}
\volume{6}
\issue{2}
\pubyear{2012}
\firstpage{753}
\lastpage{771}

\makeatletter
\newcolumntype{d}[1]{D{.}{.}{#1}}
\newcommand{\notag}{\nonumber}
\newcommand{\bs}{\mathbf}
\newcommand{\by}{\mathbf{y}}
\newcommand{\bZ}{\mathbf{Z}}
\newcommand{\bbeta}{{\boldsymbol{\beta}}}
\newcommand{\bgamma}{{\boldsymbol{\gamma}}}
\newcommand{\btheta}{{\boldsymbol{\theta}}}

\makeatother

\begin{document}
\begin{frontmatter}

\title{Bayesian modeling longitudinal dyadic data with nonignorable
dropout, with application to a~breast cancer study}
\runtitle{Longitudinal dyadic data with nonignorable dropout}

\begin{aug}
\author[A]{\fnms{Guangyu} \snm{Zhang}\ead[label=e1]{guangyuz@umd.edu}}
\and
\author[B]{\fnms{Ying} \snm{Yuan}\corref{}\thanksref{au1}\ead[label=e2]{yyuan@mdanderson.org}}
\runauthor{G. Zhang and Y. Yuan}
\affiliation{University of Maryland and  University~of~Texas~MD~Anderson~Cancer~Center}
\address[A]{Department of Epidemiology\\ \quad  and Biostatistics\\
University of Maryland\\
College Park, Maryland 20742 \\ USA \\
\printead{e1}} %adresu isvedimo komanda gale!
\address[B]{Department of Biostatistics\\
 University of Texas\\ \quad  MD Anderson Cancer Center\\
Houston, Texas 77030 \\ USA \\
\printead{e2}}
\end{aug}
\thankstext{au1}{Supported in part by
National Cancer Institute Grant R01CA154591.}

% HISTORY:
\received{\smonth{1} \syear{2011}}
\revised{\smonth{9} \syear{2011}}

% ABSTRACT
%
\begin{abstract}
Dyadic data are common in the social and behavioral sciences, in which
members of dyads are correlated due to the interdependence structure
within dyads. The analysis of longitudinal dyadic data becomes complex
when nonignorable dropouts occur. We propose a~fully Bayesian
selection-model-based approach to analyze longitudinal dyadic data with
nonignorable dropouts. We model repeated measures on subjects by a
transition model and account for within-dyad correlations by random
effects. In the model, we allow subject's outcome to depend on his/her
own characteristics and measure history, as well as those of the other
member in the dyad. We further account for the nonignorable missing
data mechanism using a selection model in which the probability of
dropout depends on the missing outcome. We propose a Gibbs sampler
algorithm to fit the model. Simulation studies show that the proposed
method effectively addresses the problem of nonignorable dropouts. We
illustrate our methodology using a longitudinal breast cancer study.
\end{abstract}

% KEYWORDS
%

\end{frontmatter}

%s1 ###
%s1 #&#
\section{Introduction}\label{sec1}\label{secintro}
Dyadic data are common in psychosocial and behavioral studies [\citet{KenKasCoo06}]. Many social phenomena, such as dating and
marital relationships, are interpersonal by definition, and, as a
result, related observations do not
refer to a single person but rather to both persons involved in the
dyadic relationship. Members of dyads often
influence each other's cognitions, emotions and behaviors, which leads to
interdependence in a relationship. For example,
a husband's (or wife's) drinking behavior may lead to lowered marital
satisfaction for the wife (or husband). A
consequence of interdependence is that observations of the two
individuals are correlated. For example, the marital satisfaction
scores of husbands
and wives tend to be positively correlated. One of the primary objectives
of relationship research is to understand the interdependence of
individuals within dyads and how the attributes and behaviors of one
dyad member impact the outcome of the other dyad member.

In many studies, dyadic outcomes are measured over time, resulting in
longitudinal dyadic data. Repeatedly measuring dyads brings in two
complications. First, in addition to the within-dyad correlation,
repeated measures on each subject are also correlated, that is,
within-subject correlation. When analyzing longitudinal dyadic data, it
is important to account for these two types of correlations
simultaneously; otherwise, the analysis results may be invalid. The
second complication is that longitudinal dyadic data are prone to the
missing data problem caused by dropout, whereby subjects are lost to
follow-up and their responses are not observed thereafter. In
psychosocial dyadic studies, the dropouts are often nonignorable or
informative in the sense that the dropout depends on missing values. In
the presence of the nonignorable dropouts, conventional statistical
methods may be invalid and lead to severely biased estimates [\citet{LitRub02}].

There is extensive literature on statistical modeling of nonignorable
drop\-outs in longitudinal studies.
Based on different factorizations of the likelihood of the outcome process
and the dropout process, \citet{Lit95} identified two broad classes of
likelihood-based nonignorable models: selection models [\citet{WuCar88}; \citet{DigKen94}; Follman and Wu (\citeyear{FolWu95});
 \citet{GlyLaiRub86}]
and pattern mixture
models [\citet{WuBai89};  Little (\citeyear{Lit93}, \citeyear{Lit94}); Hogan and Laird
(\citeyear{HogLai}); \citet{Roy03}; \citet{HogLinHer04}]. Other likelihood-based
approaches that do not directly belong to this classification have also
been proposed in the literature, for example, the mixed-effects hybrid
model by \citet{YuaLit09} and a class of nonignorable models by
\citet{Tsoetal10}. Another general approach
for dealing with nonignorable dropouts is based on estimation equations
and includes \citet{RobRotZha95}, \citet{RotRobSch98},
\citet{SchRotRob99} and
\citet{Far10}. Recent reviews of methods handling nonignorable
dropouts in
longitudinal data can be found in \citet{VerMol00},
\citet{MolKen07}, Little (\citeyear{Lit09}), \citet{IbrMol09} and \citet{DanHog08}. In spite of the rich body of
literature noted above, to the best of our knowledge, the nonignorable
dropout problem has not been addressed in the context of longitudinal
dyadic data. The interdependence structure within dyads brings new
challenges to this missing data problem. For example, within dyads, one
member's outcome often depends on his/her covariates, as well as the
other member's outcome and covariates. Thus, the dropout of the other
member in the dyad causes not only a missing (outcome) data problem for
that member, but also a~missing (covariate) data problem for the member
who remains in the study.\looseness=-1\vadjust{\goodbreak}

We propose a fully Bayesian approach to deal with longitudinal dyadic
data with nonignorable dropouts based on a selection model. Specifically,
we model each subject's longitudinal measurement process using a
transition model, which includes both the patient's and spouse's
characteristics as covariates in order to capture the interdependence
between patients and their spouses. We account for the within-dyad
correlation by introducing dyad-specific random effects into the
transition model. To accommodate the nonignorable dropouts, we take the
selection model approach by directly modeling the relationship between
the dropout process and missing outcomes using a discrete time survival model.

The remainder of the article is organized as follows. In Section~\ref{sec2} we
describe our motivating data collected from a longitudinal dyadic
breast cancer study. In Section~\ref{sec3} we propose a Bayesian
selection-model-based approach for longitudinal dyad data with
informative nonresponse, and provide estimation procedures using a
Gibbs sampler in Section~\ref{sec4}. In Section~\ref{sec5} we present simulation studies
to evaluate the performance of the proposed method.
In Section~\ref{sec6} we illustrate our method by analyzing a breast cancer data
set and we provide conclusions in Section~\ref{sec7}.

%s2 ###
%s2 #&#
\section{A motivating example}\label{sec2}
Our research is motivated by a single-arm dyadic study focusing on
physiological and psychosocial aspects of pain among patients with
breast cancer and their spouses [\citet{Badetal10}].
For individuals with breast cancer, spouses are most commonly reported
as being the primary sources of support [\citet{Kiletal98}], and spousal support is associated with lower
emotional distress and depressive symptoms in these patients [\citet{Robetal94}]. One specific aim of the study is to
characterize the depression experience due to metastatic breast cancer
from both patients' and spouses' perspectives, and examine the dyadic
interaction and interdependence of patients and spouses over time
regarding their depression. The results will be used to guide the
design of an efficient prevention program to decrease depression among
patients. For example, conventional prevention programs typically apply
interventions to patients directly. However, if we find that the
patient's depression depends on both her own and spouse's previous
depression history and chronic pain, when designing a prevention
program to improve the depression management and pain relief, we may
achieve better outcomes by targeting both patients and spouses
simultaneously rather than targeting patients only. In this study,
{female patients who had initiated metastatic breast cancer treatment
were approached by the project staff. Patients meeting the eligibility
criteria (e.g., speak English, experience pain due to the breast
cancer, having a male spouse or significant other, be able to carry on
pre-disease performance, be able to provide informed consent) were
asked to participate the study on a voluntary basis. The participation
of the study would not affect their treatment in any way.

Depression in patients and spouse was measured at three time points
(baseline, 3 months and 6 months) using the Center for Epidemiologic
Studies Depression Scale (CESD) questionnaires. However, a substantial
number of dropouts occurred. Baseline CESD measurements were collected
from 191 couples; however, at 3 months, 101 couples (105 patients and
107 spouses) completed questionnaires, and at 6 months, 73 couples (76
patients and 79 spouses) completed questionnaires. The missingness of
the CESD measurements is likely related to the current depression
levels of the patients or spouses, thus an nonignorable missing data
mechanism is assumed for this study. Consequently, it is important to
account for the nonignorable dropouts in this data analysis; otherwise,
the results may be biased, as we will show in Section~\ref{sec6}.

%s3 ###
%s3 #&#
\section{Models}\label{sec3}

Consider a longitudinal dyadic study designed to collect $J$
repeated measurements of a response $Y$ and a vector of covariates $\bs{X}$ for each
of $n$ dyads. Let $Y_{kij}$, $\bs{X}_{kij}$ and $\bs
{H}_{kij}=(y_{ki,j-1}, \ldots, y_{ki1})^T$ denote the outcome,
$p\times1$ covariate vector and outcome history, respectively, for the
member $k$ of dyad $i$ at the $j$th measurement time with $k=1, 2;
i=1,\ldots, n; j=1,\ldots, J$. We assume that $\bs{X}$ is fully observed
(e.g., is external or fixed by study design), but $Y$ is subject to
missingness due to dropout. The random variable~$D_{ki}$, taking values
from $2$
to $J+1$, indicates the time the member $k$ of the $i$th dyad drops
out, where
$D_{ki} = J+1$ if the subject completes the study, and $D_{ki} = j$ if
the subject
drops out between the $(j -1)$th and $j$th measurement time,
that is, $\{y_{ki1}, \ldots , y_{ki,j-1}\}$ are observed and
$\{y_{kij}, \ldots , y_{kiJ}\}$ are missing. We assume
at least 1 observation for each subject, as subjects without any
observations have no information and are often excluded from the analysis.

When modeling longitudinal dyadic data, we need to consider two types
of correlations: the within-subject correlation due to repeated
measures on a subject, and the within-dyad correlation due to the
dyadic structure. We account for the first type of correlation by a
transition model, and the second type of correlation by dyad-specific
random effects $b_i$, as follows:
%e3.3 ###
%e3.2 ###
%e3.1 ###
%
%e3.1 #&#
%e3.2 #&#
\begin{eqnarray}\label{transition}
Y_{1ij}|b_i &=& b_{i} + \alpha_1+ \bs{H}^{T}_{1ij}{\bbeta}_{1} + \bs
{H}^{T}_{2ij}{\bgamma}_{1} + \bs{X}_{1ij}^{T}\tilde{\bbeta}_{1} +
\bs{X}_{2ij}^{T}\tilde{\bgamma}_{1} + e_{1ij}, \notag\\
Y_{2ij}|b_i &=& b_{i} + \alpha_2 + \bs{H}^{T}_{2ij}{\bbeta}_{2} +
\bs{H}^{T}_{1ij}{\bgamma}_{2} + \bs{X}_{2ij}^{T}\tilde{\bbeta}_{2}
+ \bs{X}_{1ij}^{T}\tilde{\bgamma}_{2} + e_{2ij},
\\
b_{i} &\sim& N(0, \tau^2_b).\notag
\end{eqnarray}
Regression parameters in this random-effects transition model have
intuitive interpretations similar to those of the actor--partner
interdependence model, a conceptual framework proposed by \citet{CooKen05} to study dyadic relationships in the social sciences and
behavior research fields. Specifically,~$\tilde\bbeta_{1}$ and
${\bbeta}_{1}$ represent the ``actor'' effects of the patient, which
indicate how the covariates\vadjust{\goodbreak} and the outcome history of the patient
(i.e., $\bs{X}_{1ij}$ and $\bs{H}_{1ij}$) affect her own current
outcome, whereas $\tilde\bgamma_{1}$ and ${\bgamma}_{1}$ represent
the ``partner'' effects for
the patient, which indicate how the covariates and the outcome history
of the spouse (i.e., $\bs{X}_{2ij}$ and $\bs{H}_{2ij}$) affect the
outcome of the patient. Similarly, $\tilde\bbeta_{2}$ and ${\bbeta
}_{2}$ characterize the actor effects and $\tilde{\bgamma}_{2}$ and
$\bgamma_2$ characterize the partner effects for the spouse of the
patient. Estimates of the actor and partner effects provide important
information about the interdependence within dyads.
We assume that residuals $e_{1ij}$ and $e_{2ij}$ are independent and follow
normal distributions $N(0, \sigma_1^2)$ and $N(0, \sigma_2^2)$,
respectively; and $e_{1ij}$ and $e_{2ij}$ are independent of random
effects $b_i$'s. The parameters $\alpha_1$ and $\alpha_2$ are
intercepts for the patients and spouses, respectively.

In many situations, the conditional distribution of $Y_{kij}$ given
$\bs{H}_{kij}$ and~$\bs{X}_{kij}$ depends only on the $q$ prior
outcomes $y_{ki,j-1}, \ldots, y_{ki,j-q}$ and $\bs{X}_{kij}$. If this
is the case, we obtain the so-called $q$th-order transition model, a
type  of transition model that is most useful in practice [\citet{Digetal02}]. The choice of the model order $q$
depends on subject matters. In many applications, it is often
reasonable to set $q=1$ when the current outcome depends on only the
last observed previous outcome, leading to commonly used Markov models.
The likelihood ratio test can be used to assess whether a specific
value of~$q$ is appropriate [\citet{KalLaw85}].
Auto-correlation analysis of the outcome history also can provide
useful information to determine the value of $q$ [\citet{Got81};
\citet{KenOrd90}].

Define $\bs{Y}_{ki}= (Y_{ki1}, \ldots, Y_{kid_{ki}})$ and $\bs
{X}_{ki}= (\bs{X}_{ki1}, \ldots, \bs{X}_{kid_{ki}} ) $ for $k=1, 2$.
Given \{$\bs{X}_{1i}, \bs{X}_{2i} \}$ and the random effect $b_i$,
the joint log likelihood of $(\bs{Y}_{1i}, \bs{Y}_{2i})$ for the
$i$th dyad under the $q$th-order (random-effects) transition model is
given by
\begin{eqnarray*}
&&\hspace*{-2pt} \ell_i(\bs{Y}_{1i}, \bs{Y}_{2i}|\bs{X}_{1i}, \bs{X}_{2i}, b_i)
\\
&&\hspace*{-2pt} \qquad  = \sum_{j=q+1}^{d_{1i}} \ell_{ij}(Y_{1ij}|\bs{X}_{1ij}, \bs
{X}_{2ij}, \bs{H}_{1ij}, \bs{H}_{2ij}, b_i ) + \ell_i(Y_{1i1},
\ldots, Y_{1iq}|\bs{X}_{1i}, \bs{X}_{2i}) \\
&&\hspace*{-2pt} \qquad  \quad {}+\sum_{j=q+1}^{d_{2i}} \ell_{ij}(Y_{2ij}| \bs{X}_{1ij}, \bs
{X}_{2ij}, \bs{H}_{1ij}, \bs{H}_{2ij}, b_i ) + \ell_i(Y_{2i1},
\ldots, Y_{2iq}|\bs{X}_{1i}, \bs{X}_{2i}),
\end{eqnarray*}
where $\ell_{ij}(Y_{kij}|\bs{X}_{1ij}, \bs{X}_{2ij}, \bs{H}_{1ij},
\bs{H}_{2ij}, b_i )$ is the likelihood corresponding to model (\ref
{transition}), and $\ell_i(Y_{ki1}, \ldots, Y_{1iq}|\bs{X}_{1i}, \bs
{X}_{2i})$ is assumed free of $\boldsymbol{\eta}_k = (\alpha_{k}, \bbeta_k,
\tilde\bbeta_k,\break \bgamma_k, \tilde\bgamma_k)$, for $k=1, 2$.

An important feature of model (\ref{transition}) that distinguishes it
from the standard transition model is that the current value of the
outcome $Y$ depends on not only the subject's outcome history, but also
the spouse's outcome history. Such a ``partner'' effect is of particular
interest in dyadic studies because it reflects the interdependence
between the patients and spouses. This interdependence within dyads
also makes the missing data problem more challenging. Consider a dyad
consisting of subjects $A$ and~$B$ and that~$B$ drops out prematurely.
Because the outcome history of $B$ is used as a~covariate in the
transition model of $A$, when $B$ drops out, we face not only the
missing outcome (for $B$) but also missing covariates (for $A$). We
address this dual missing data problem using the data augmentation
approach, as described in Section~\ref{sec4}.

To account for nonignorable dropouts, we employ the discrete time
survival model [\citet{Agr02}] to jointly model the missing data
mechanism. Specifically, we assume that the distribution of $D_{ki}$
depends on both the past history of the longitudinal process and the
current outcome $Y_{kij}$, but not on future observations. Define the
discrete hazard rate $\lambda_{kij}(\bs{H}_{kij}, Y_{kij}, \bs
{X}_{kij}) = \operatorname{Pr}(D_{ki}=j|D_{ki}>j-1, \bs{H}_{kij}, Y_{kij}, \bs
{X}_{kij})$. It follows that the probability of dropout for the member
$k$ in the $i$th dyad is given by
\begin{eqnarray*}
&&\hspace*{-5pt}\operatorname{Pr}(D_{ki}=d | \bs{H}_{kij}, Y_{kij}, \bs{X}_{kij}) \\
&&\hspace*{-5pt} \quad = \cases{
\displaystyle  \prod_{j=2}^{d-1} \{1-\lambda_{kij}(\bs{H}_{kij},
Y_{kij}, \bs{X}_{kij})\}\lambda_{kid} (\bs{H}_{kid}, Y_{kid}, \bs
{X}_{kid}),& \quad  if    $d\leq J$, \cr \displaystyle \prod_{j=2}^J \{1-\lambda
_{kij}(\bs{H}_{kij}, Y_{kij}\bs{X}_{kij}) \},& \quad if     $d=J+1$.
}
\end{eqnarray*}

We specify the discrete hazard rate $\lambda_{kij}(\bs{H}_{kij},
Y_{kij}, \bs{X}_{kij})$ using the logistic regression model:
%e3.6 ###
%e3.5 ###
%e3.4 ###
%
%e3.3 #&#
\begin{eqnarray}\label{dropmodel}
 \qquad \operatorname{Logit}(\lambda_{1ij}(\bs{H}_{1ij}, Y_{1ij}, \bs{X}_{1ij})) &=&
c_i + \xi_1+\bs{X}_{1ij}^{T}\boldsymbol{\psi}_1 + \bs{H}_{1ij}^{T}\boldsymbol
{\delta}_1 + \phi_1 Y_{1ij}, \notag\\
 \quad \operatorname{Logit}(\lambda_{2ij}(\bs{H}_{2ij}, Y_{2ij}, \bs{X}_{2ij})) &=&
c_i + \xi_2 + \bs{X}_{2ij}^{T}\boldsymbol{\psi}_2 + \bs{H}_{2ij}^{T}\boldsymbol
{\delta}_2 + \phi_2 Y_{2ij},  \\
c_i  &\sim&  N(0, \tau^2_c),\notag
\end{eqnarray}
where $c_i$ is the random effect accounting for the within-dyadic
correlation, and $\xi_k, \boldsymbol{\psi}_k, \boldsymbol{\delta}_k$ and $\phi_k,
k=1, 2,$ are unknown parameters. In this dropout model, we assume that,
conditioning on the random effects, a subject's covariates, past
history and current (unobserved) outcome, the dropout probability of
this subject is independent of the characteristics and outcomes of the
other member in the dyad. The spouse may indirectly affect the dropout
rate of the patient through influencing the patient's depression
status; however, when conditional on the patient's depression score,
the dropout of the patient does not depend on her spouse's depression score.

In practice, we often expect that, given $Y_{kij}$ and $Y_{ki,j-1}$,
the conditional dependence
of $D_{ki}$ on $Y_{ki,j-2},\ldots, Y_{ki,1}$ will be negligible
because, temporally, the patient's (current) decision of dropout is
mostly driven by his (or her) current and the most recent outcome
statuses. Using the breast cancer study as an example, we do not expect
that the early history of depression plays an important role for the
patient's current decision of dropout; instead, the patient drops out
typically because she is currently experiencing or most recently
experienced high depression. The early history may influence the
dropout but mainly through its effects on the current depression
status. Once conditioning on the current and the most recent depression
statuses, the influence from the early history is essentially
negligible. Thus, we use a~simpler form of the discrete hazard model
%
%e3.4 #&#
\begin{eqnarray}
\operatorname{Logit}(\lambda_{kij}(\bs{H}_{kij}, Y_{kij}, \bs{X}_{kij})) =
c_i + \xi_k + \bs{X}_{kij}^{T}\boldsymbol{\psi}_k + {\delta}_k Y_{ki, j-1}
+ \phi_k Y_{kij},\nonumber\\
\eqntext{ k=1, 2.}
\end{eqnarray}

%s4 ###
%s4 #&#
\section{Estimation}\label{sec4}\label{secEstimation}
Under the Bayesian paradigm, we assign the following vague priors to
the unknown parameters and fit the proposed model using a Gibbs sampler:
\begin{eqnarray*}
\alpha_{k}, \bbeta_k, \tilde{\bbeta}_k, \bgamma_k, \tilde{\bgamma
}_k, \xi_k, \boldsymbol{\psi}_k, {\delta}_k \mbox{ and }   \phi_k &\sim&
\mathrm{constant}, \qquad  k=1, 2;\\
\sigma_k^2 &\sim& \mathit{IG}(a, b), \qquad  k=1, 2; \\
\tau_b^2 &\sim& \mathit{IG}(a, b);\\
\tau_c^2 &\sim& \mathit{IG}(a, b);
\end{eqnarray*}
where $\mathit{IG}(a,b)$ denote an inverse gamma distribution with a shape
parameter $a$ and a scale parameter $b$. We set $a$ and $b$ at smaller
values, such as 0.1, so that the data dominate the prior information.
Let $\mathbf{y}_\mathrm{obs}$ and $\mathbf{y}_\mathrm{mis}$ denote
the observed and missing part of the data, respectively. Considering
the $k$th iteration of the Gibbs sampler, the first step of the
iteration is ``data augmentation'' [\citet{TanWon87}], in which
the missing data~$\mathbf{y}_\mathrm{mis}$ are generated from their
full conditional distributions. Without loss of generality, suppose for
the $i$th dyad, member 2 drops out of the study no later than member 1,
that is, $d_{1i} \ge d_{2i}$, and let $d_i =\operatorname{max}(d_{1i}, d_{2i})$.
Assuming a~first-order ($q = 1$) transition model (or Markov model) and
letting $\btheta$ denote a~generic symbol that represents the values
of all other model parameters, the data augmentation consists of the
following 3 steps:
\begin{longlist}[(3)]
\item[(1)] For $j=d_{2i}, \ldots, d_i-1$, draw $y_{2ij}$ from the
conditional distribution
\begin{eqnarray*}
y_{2ij}|\mathbf{y}_\mathrm{obs}, \btheta&\propto& N \biggl( \frac
{\sigma_2^{-2} \mu_1^* + \beta_2\sigma_2^{-2} \mu_2^* + \gamma
_1\sigma_1^{-2}\mu_3^*}
{\sigma_2^{-2} + \beta_2^2\sigma_2^{-2} +\gamma_1^2\sigma_1^{-2}},
\frac{1}{\sigma_2^{-2} + \beta_2^2\sigma_2^{-2} +\gamma_1^2\sigma
_1^{-2}}  \biggr)\\
&&{} \times\lambda_{2id_{2i}} (\bs{H}_{2id_{2i}}, y_{2id_{2i}}, \bs
{X}_{2id_{2i}})^{I(j=d_{2i})},
\end{eqnarray*}
where
\begin{eqnarray*}
\mu_1^* &=& b_{i} +{\beta}_{2}y_{2i,j-1} + {\gamma}_{2}y_{1i,j-1} +
\alpha_{2}+\bs{X}_{2ij}^{T}\tilde\bbeta_{2} + \bs
{X}_{1ij}^{T}\tilde\bgamma_{2}, \\
\mu_2^* &=& y_{2i, j+1} - b_i -\gamma_2 y_{1ij} -\alpha_{2}-\bs
{X}_{2i, j+1}^{T}\tilde\bbeta_{2} - \bs{X}_{1i, j+1}^{T}\tilde
\bgamma_{2}, \\
\mu_3^* &=& y_{1i, j+1}-b_i-\beta_1 y_{1ij}-\alpha_{1}- \bs{X}_{1i,
j+1}^{T}\tilde\bbeta_{1} - \bs{X}_{2i, j+1}^{T}\tilde\bgamma_{1}.
\end{eqnarray*}
\item[(2)] Draw $y_{2i,d_i}$ from the conditional distribution
\[
y_{2i,d_i}|\mathbf{y}_\mathrm{obs}, \btheta\sim N(b_i + y_{2i,
d_i-1}{\beta}_{2} + y_{1i,d_i-1}{\gamma}_{2}+ \alpha_{2} + \bs
{X}_{2id_i}^{T}\tilde\bbeta_{2} + \bs{X}_{1id_i}^{T}\tilde\bgamma
_{2}, \sigma_2^2).
\]

\item[(3)] Draw $y_{1i, d_i}$ from the conditional distribution
\begin{eqnarray*}
y_{1i,d_i}|\mathbf{y}_\mathrm{obs}, \btheta&\propto& N( b_{i} +
y_{1i, d_i-1}{\beta}_{1} + y_{2i, d_i-1}{\gamma}_{1} + \alpha
_{1}+\bs{X}_{1id_i}^{T}\tilde\bbeta_{1} + \bs{X}_{2id_i}^{T}\tilde
\bgamma_{1}, \sigma_1^2)\\
&& \times\lambda_{1id_{1i}} (\bs{H}_{1id_{i}}, y_{1id_{i}}, \bs
{X}_{1id_{i}}).
\end{eqnarray*}
\end{longlist}

Now, with the augmented complete data $\by=\{\by_\mathrm{obs}, \by_\mathrm{mis}\}$, the parameters are drawn alternatively as follows:
\begin{longlist}[(10)]
\item[(4)] For $i=1, \ldots, n$, draw random effects $b_i$ from the
conditional distribution
\begin{eqnarray*}
b_i |\by, \btheta
&=& N \biggl(\frac{\sum_{j=2}^{d_i} (y_{1ij}-\mu_{1ij})\sigma_2^2
\tau_b^2 + \sum_{j=2}^{d_i}(y_{2ij}-\mu_{2ij}) \sigma_1^2 \tau_b^2}
{(d_i-1)\sigma_1^2\tau_b^2+ (d_i-1)\sigma_2^2\tau_b^2 + \sigma_1^2
\sigma_2^2},
\\
&& \hphantom{ N \biggl(} \frac{\sigma_1^2 \sigma_2^2 \tau_b^2 }{(d_i-1)\sigma
_1^2\tau_b^2+ (d_i-1)\sigma_2^2\tau_b^2 + \sigma_1^2 \sigma
_2^2} \biggr),
\end{eqnarray*}
where
\begin{eqnarray*}
\mu_{1ij} &=& y_{1i, j-1} \beta_1 + y_{2i, j-1} \gamma_1 + \alpha
_{1}+\bs{X}_{1ij}^T \tilde\bbeta_1 + \bs{X}_{2ij}^T \tilde\bgamma
_1 ,\\
\mu_{2ij} &=& y_{2i, j-1} \beta_2 + y_{1i, j-1} \gamma_2 + \alpha
_{2}+\bs{X}_{2ij}^T \tilde\bbeta_2 + \bs{X}_{1ij}^T \tilde\bgamma_2.
\end{eqnarray*}

\item[(5)] Draw $\sigma_k^2$ from the conditional distribution
\[
\sigma_k^2 |\by, \btheta= \mathit{IG} \biggl(a + \frac{\sum
_{i=1}^n(d_i-1)}{2} , b+\frac{\sum_{i=1}^n \sum
_{j=2}^{d_i}(y_{kij}-u_{kij})^2}{2} \biggr),
\]
where
\begin{eqnarray*}
u_{1ij} &=& b_i + y_{1i, j-1} \beta_1 + y_{2i, j-1} \gamma_1 + \alpha
_{1}+ \bs{X}_{1ij}^T \tilde\bbeta_1 + \bs{X}_{2ij}^T \tilde\bgamma
_1, \\
u_{2ij} &=& b_i + y_{2i, j-1} \beta_2 + y_{1i, j-1} \gamma_2 + \alpha
_{2} +\bs{X}_{2ij}^T \tilde\bbeta_2 + \bs{X}_{1ij}^T \tilde\bgamma_2.
\end{eqnarray*}

\item[(6)] Draw $\tau_b^2$ from the conditional distribution
\[
\tau_b^2|\by, \btheta= \mathit{IG} \biggl(a + \frac{n}{2}, b+\frac{\sum
_{i=1}^n b_i^2}{2} \biggr).
\]

\item[(7)] Draw $ \boldsymbol{\eta}_1=(\alpha_1, \beta_1, \gamma_1, \tilde
\bbeta_1, \tilde\bgamma_1)$ from the normal distribution
\[
\boldsymbol{\eta}_1 |\by, \btheta= N\bigl((\bZ_1^T \bZ_1)^{-1} {\bZ_1^T} (\by
_1-b_i), (\bZ_1^T \bZ_1)^{-1}\sigma_1^2\bigr),
\]
where
$\bs{y}_1=(y_{11, 2}, \ldots, y_{11,d_1}, \ldots,y_{1i,2}, \ldots,
y_{1i,d_i}, \ldots, y_{1n,2}, \ldots, y_{1n,d_n})^T$ and
\[
\bs{Z}_1 =  \left(
\matrix{
1& \cdots& 1 & \cdots& 1 & \cdots& 1 & \cdots\cr
y_{11,1} & \cdots& y_{11,d_i-1} & \cdots& y_{1i,1} & \cdots&
y_{1i,d_i-1} & \cdots\cr
y_{21,1} & \cdots& y_{21,d_i-1} & \cdots& y_{2i,1} & \cdots&
y_{2i,d_i-1} & \cdots\cr
\bs{X}_{11,2} & \cdots& \bs{X}_{11,d_1} & \cdots& \bs{X}_{1i,2} &
\cdots& \bs{X}_{1i,d_i} & \cdots\cr
\bs{X}_{21,2} & \cdots& \bs{X}_{21,d_1} & \cdots& \bs{X}_{2i,2} &
\cdots& \bs{X}_{2i,d_i} & \cdots
}\right)^T.
\]

\item[(8)] Similarly, draw $ \boldsymbol{\eta}_2 =(\alpha_2, \beta_2, \gamma_2,
\tilde\bbeta_2, \tilde\bgamma_2)$ from the conditional distribution
\[
\boldsymbol{\eta}_2 | \bs{y}, \btheta= N\bigl((\bZ_2^T \bZ_2)^{-1} {\bZ_2^T}
(\by_2-b_i), (\bZ_2^T \bZ_2)^{-1}\sigma_2^2\bigr),
\]
where
$\bZ_2$ and $\bs{y}_2$ are defined in a similar way to $\bZ_1$ and
$\bs{y}_1.$

\item[(9)] Draw $\boldsymbol{\varpi}_1=(\xi_1, \boldsymbol{\psi}_1 ,\delta_1,\phi_1)$
and $\boldsymbol{\varpi}_2=(\xi_2, \boldsymbol{\psi}_2,\delta_2,\phi_2)$ from
the conditional distributions
\begin{eqnarray*}
\boldsymbol{\varpi}_1 |\by, \btheta&\propto&\prod_{i=1}^{n}\prod
_{j=2}^{d_{1i}-1} (1-\lambda_{1ij}) \lambda_{1id_{1i}}, \\
\boldsymbol{\varpi}_2 |\by, \btheta&\propto&\prod_{i=1}^{n}\prod
_{j=2}^{d_{2i}-1} (1-\lambda_{2ij}) \lambda_{2id_{2i}}.
\end{eqnarray*}

\item[(10)] Draw random effects $c_i$ from the conditional distribution
\begin{eqnarray*}
c_i | \by, \btheta
\propto N(0, \tau^2_c) \prod_{j=2}^{d_{1i}-1} (1-\lambda_{1ij})
\lambda_{1id_{1i}}\prod_{j=2}^{d_{2i}-1} (1-\lambda_{2ij}) \lambda
_{2id_{2i}}.
\end{eqnarray*}

\item[(11)] Draw $\tau_c^2$ from the conditional distribution
\[
\tau_c^2 | \by, \btheta= \mathit{IG} \biggl(a + \frac{n}{2}, b+\frac{\sum
_{i=1}^n c_i^2}{2} \biggr).
\]
\end{longlist}

%s5 ###
%s5 #&#
\section{Simulation studies}\label{sec5}\label{secsimu}

We conducted two simulation studies (A and B). Simulation A consists of
500 data sets, each with 200 dyads and three repeated measures.
For the $i$th dyad, we generated the first measurements,~$Y_{1i1}$ and
$Y_{2i1}$, from normal distributions $N(5,1)$ and $N(7,1)$,
respectively, and generated the second and third measurements based on
the following random-effects transition model:
\begin{eqnarray*}
Y_{1ij}| b_i &\sim& N(b_i + \beta_1Y_{1i, j-1}+ \gamma_1Y_{2i, j-1}+
\tilde{\beta}_1X_1 + \tilde{\gamma}_1X_2,1),  \qquad  j=2, 3, \\
Y_{2ij}| b_i &\sim& N(b_i + \beta_2Y_{2i, j-1}+ \gamma_2Y_{1i, j-1}+
\tilde{\beta}_2X_2 + \tilde{\gamma}_2X_1,1),  \qquad  j=2, 3, \\
b_i &\sim& N(0, 1),
\end{eqnarray*}
where $\beta_1=\gamma_1=0.5$, $\beta_2=\gamma_2=0.6$, $\tilde
{\beta}_1=\tilde{\gamma}_1=\tilde{\beta}_2=\tilde{\gamma}_2=1$,
and covariates~$X_1$ and $X_2$ were generated independently from $N(0,
1)$. We assumed that the baseline (first) measurements $Y_{1i1}$ and
$Y_{2i1}$ were observed for all subjects, and the hazard of dropout at
the second and third measurement times depended on the current and last
observed values of $Y$, that is,
\begin{eqnarray*}
\operatorname{logit}(\lambda_{1ij}|c_i)&=&c_i-Y_{1ij}-0.5Y_{1i,j-1}-6,
\qquad
j=2, 3, \\
\operatorname{logit}(\lambda_{2ij}|c_i)&= &
c_i-Y_{2ij}-0.5Y_{2i,j-1}-6,\qquad
j=2, 3, \\
c_i  &\sim&  N(0,1).
\end{eqnarray*}
Under this dropout model, on average, 24\% (12\% of member 1 and 13\%
of member 2) of the dyads dropped out at the second time point and 45\%
(26\% of member~1 and 30\% of member~2) dropped out at the third
measurement time. We applied the proposed method to the simulated data
sets. We used 1,000 iterations to burn in and made inference based on
10,000 posterior draws. For comparison purposes, we also conducted
complete-case and available-case analyses. The complete-case analysis
was based on the data from dyads who completed the follow-up, and the
available-case analysis was based on all observed data (without
considering the missing data mechanism).

%t1 ###
%
%t1 #&#
\begin{table}
\tabcolsep=0pt
\caption{Bias, standard error (SE) and coverage rate of 95\% credible
intervals under different methods for simulation A}\label{tbTransModel1}
\label{tab1}\begin{tabular*}{\textwidth}{@{\extracolsep{4in minus 4in}}ld{2.2}ccd{2.2}ccd{2.2}cc@{}}
\hline
& \multicolumn{3}{c}{\textbf{Complete-case analysis}} & \multicolumn{3}{c}{\textbf{Available-case analysis}}
& \multicolumn{3}{c@{}}{\textbf{Proposed method}}\\[-5pt]
& \multicolumn{3}{c}{\hrulefill} & \multicolumn{3}{c}{\hrulefill} & \multicolumn{3}{c@{}}{\hrulefill}\\
\textbf{Parameter} & \multicolumn{1}{c}{\textbf{Bias}} & \textbf{SE} & \textbf{Coverage}  & \multicolumn{1}{c}{\textbf{Bias}}
& \textbf{SE} & \textbf{Coverage} & \multicolumn{1}{c}{\textbf{Bias}} & \textbf{SE} & \textbf{Coverage} \\ \hline
$\beta_1$ & -0.03 & 0.06 & 0.93 & -0.01 & 0.05 & 0.94& -0.01 &
0.05& 0.95 \\
$\gamma_1$ & -0.06 & 0.05 & 0.81 & -0.03 & 0.04 & 0.88 & 0.07 &
0.04& 0.96 \\
$\tilde{\beta}_1$ & -0.16 & 0.12 & 0.72 & -0.10 & 0.10 & 0.81 &
0.05 & 0.08& 0.94 \\
$\tilde{\gamma}_1$ & -0.17 & 0.12 & 0.75 & -0.10 & 0.10 & 0.78 &
0.02 & 0.09& 0.97 \\[4pt]
$\beta_2$ & -0.06 & 0.06 & 0.89 & -0.06 & 0.05 & 0.84 & 0.08 & 0.05&
0.97 \\
$\gamma_2$ & -0.04 & 0.05 & 0.87 & -0.00 & 0.04 & 0.95 & -0.04 &
0.06& 0.96 \\
$\tilde{\beta}_2$ & -0.17 & 0.12 & 0.73 & -0.10 & 0.10 & 0.84&-0.01
& 0.12& 0.95 \\
$\tilde{\gamma}_2$& -0.17 & 0.12 & 0.72 & -0.10 & 0.10 & 0.81 &
0.01 & 0.09 & 0.97 \\
\hline
\end{tabular*}
\end{table}

Table \ref{tbTransModel1} shows the bias, standard error (SE) and
coverage rate of the 95\% credible interval (CI) under different
approaches. We can see that the proposed method substantially
outperformed the complete-case and available-case analyses. Our method
yielded estimates with smaller bias and coverage rates close to the
95\% nominal level. In contrast, the complete-case and available-case
analyses often led to larger bias and poor coverage rates. For example,
the bias of the estimate of $\tilde{\beta}_1$ under the complete-case
and available-case analyses were $-0.16$ and $-0.10$, respectively,
substantially larger than that under the proposed method (i.e., 0.05);
the coverage rate using the proposed method was about 94\%, whereas
those using the complete-case and available-case analyses were under
82\%.

The second simulation study (Simulation B) was designed to evaluate the
performance of the proposed method when the nonignorable missing data
mechanism is misspecified, for example, data actually are missing at
random (MAR). We generated the first measurements, $Y_{1i1}$ and
$Y_{2i1}$, from normal distribution $N(3,1)$ independently, and
generated the second and third measurements based on the same
transition model as in Simulation A. We assumed the hazard\vadjust{\goodbreak} of dropout
at the second and third measurement times depended on the previous
(observed) value of $Y$ quadratically, but not on the current (missing)
value of $Y$, that is,
%e5.3 ###
%e5.2 ###
%e5.1 ###
%
%e5.1 #&#
\begin{eqnarray}\label{simu2}
\operatorname{logit}(\lambda_{1ij}|c_i)&=&c_i+Y_{1i,j-1}^2-15,  \qquad  j=2,
3,
\notag \\
\operatorname{logit}(\lambda_{2ij}|c_i)&= & c_i+Y_{2i,j-1}^2-15,  \qquad  j=2,
3,
 \\
c_i  &\sim&  N(0,1).\notag
\end{eqnarray}
Under this MAR dropout model, on average, 37\% (21\% of member 1 and
21\% of member 2) of the dyads dropped out at the second time point and
27\% (24\% of member 1 and 33\% of member 2) dropped out at the third
measurement time.
To fit the simulated data, we considered two nonignorable models with
different specifications of the dropout (or selection) model. The first
nonignorable model assumed a flexible dropout model
\[
\operatorname{logit} (\lambda_{kij}|b_i) = c_i + \xi_k+\delta_k Y_{ki,j-1}^2
+ \phi_k Y_{ki,j},
\]
which included the true dropout process (\ref{simu2}) as a specific
case with $\phi_k=0$; and the second nonignorable model took a
misspecified dropout model of the form
\[
\operatorname{logit} (\lambda_{kij}|b_i) = c_i + \xi_k+\delta_k Y_{ki,j-1}
+\phi_k Y_{ki,j}.
\]

Table \ref{SIMULATIONc} shows the bias, standard error and coverage
rate of the 95\% CI under different approaches. When the missing data
were MAR, the complete-case analysis was invalid and led to biased
estimates and poor coverage rates because the complete cases are not
random samples from the original population. In contrast, the
available-case analysis yielded unbiased estimates and coverage rates
close to the 95\% nominal level. For the nonignorable models, the one
with the flexible dropout model yielded unbiased estimates and
reasonable coverage rates, whereas the model with the misspecified
dropout model led to biased estimates (e.g., $\hat{{\beta}}_1 $ and $
\hat{{\beta}}_2 $) and poor coverage rates. This result is not
surprising because it is well known that selection models are sensitive
to the misspecification of the dropout model [\citet{LitRub02};
\citet{DanHog00}]. For nonignorable missing data, the
difficulty is that we cannot judge whether a specific dropout model is
misspecified or not based solely on observed data because the observed
data contain no information about the (nonignorable) missing data
mechanism. To address this difficulty, one possible approach is to
specify a flexible dropout model to decrease the chance of model
misspecification. Alternatively, maybe a better approach is to conduct
sensitivity analysis to evaluate how the results vary when the dropout
model varies. We will illustrate the latter approach in the next section.

%t2 ###
%
\begin{sidewaystable}
\tabcolsep=0pt
\tablewidth=\textwidth
\caption{Bias, standard error (SE) and coverage rate of 95\% credible
intervals under different methods for simulation B}
\label{SIMULATIONc}
\label{tab2}\begin{tabular*}{\textwidth}{@{\extracolsep{4in minus 4in}}ld{2.2}cccccd{2.2}ccd{2.2}cc@{}}
\hline
&&&&&&& \multicolumn{3}{c}{\textbf{Nonignorable model}}
&\multicolumn{3}{c@{}}{\textbf{Nonignorable model}} \\
& \multicolumn{3}{c}{\textbf{Complete-case analysis}} &   \multicolumn{3}{c}{\textbf{Available-case analysis}}   &\multicolumn
{3}{c}{\textbf{(flexible dropout model)}}&  \multicolumn{3}{c@{}}{\textbf{(misspecifed dropout model)}} \\[-5pt]
& \multicolumn{3}{c}{\hrulefill} &   \multicolumn{3}{c}{\hrulefill}   &\multicolumn
{3}{c}{\hrulefill}&  \multicolumn{3}{c@{}}{\hrulefill}\\
\textbf{Parameter} & \multicolumn{1}{c}{\textbf{Bias}} & \textbf{SE} & \textbf{Coverage} & \textbf{Bias} & \textbf{SE}
& \textbf{Coverage} & \multicolumn{1}{c}{\textbf{Bias}} & \textbf{SE}
& \textbf{Coverage} & \multicolumn{1}{c}{\textbf{Bias}} & \textbf{SE} & \textbf{Coverage}\\ \hline
$\beta_1$ & -0.06 & 0.08 & 0.86 & 0.00& 0.06 & 0.95& -0.01 & 0.06&
0.95 & 0.14 & 0.06& 0.78 \\
$\gamma_1$ & -0.09 & 0.08 & 0.82 & 0.00 & 0.05 & 0.96 & 0.07 & 0.05&
0.97 & -0.01 & 0.05& 0.95 \\
$\tilde{\beta}_1$ & -0.11 & 0.14 & 0.84 & 0.00 & 0.10 & 0.95 & 0.04
& 0.08& 0.96 & 0.03 & 0.08& 0.94\\
$\tilde{\gamma}_1$ & -0.13 & 0.14 & 0.84 & 0.00 & 0.10 & 0.96 &
0.02 & 0.09& 0.97 & 0.02 & 0.09& 0.98\\[4pt]
$\beta_2$ & -0.07 & 0.08 & 0.87 & 0.00 & 0.06 & 0.96 & 0.02 & 0.06&
0.97 & 0.12 & 0.06& 0.79\\
$\gamma_2$ & -0.10 & 0.08 & 0.78 & 0.00 & 0.07 & 0.96 & 0.00 & 0.06&
0.96 & -0.08 & 0.06& 0.93\\
$\tilde{\beta}_2$ & -0.14 & 0.14 & 0.82 & 0.00 & 0.10 & 0.96&0.01 &
0.12& 0.94 & 0.01 & 0.12& 0.95\\
$\tilde{\gamma}_2$& -0.14 & 0.13 & 0.83 & 0.01 & 0.10 & 0.96 &
0.01& 0.09 & 0.97& 0.01 & 0.09& 0.98 \\
\hline
\end{tabular*}
\end{sidewaystable}

%s6 ###
%s6 #&#
\section{Application}\label{sec6}\label{secapplication}

We applied our method to the longitudinal metastatic breast cancer
data. We used the first-order random-effects transition model for the
longitudinal measurement process. In the model, we included 5
covariates: chronic\vadjust{\goodbreak} pain measured by the Multidimensional Pain
Inventory (MPI) and previous CESD scores from both the patients and
spouses, and the patient's stage of cancer. In the discrete-time
dropout model, we included the subject's current and previous CESD
scores, MPI measurements and the patient's stage of cancer as
covariates. Age was excluded from the models because its estimate was
very close to 0 and not significant. We used 5,000 iterations to burn
in and made inference based on 5,000 posterior draws. We also conducted
the complete-case and available-case analyses for the purpose of comparison.

\begin{table}
\tabcolsep=0pt
\caption{Parameter estimates and 95\% credible intervals (shown in
parentheses) for the patients' and spouses' measurement models based on
the complete-case, available-case analyses and the proposed approach
for the breast cancer data}\label{tbTransModel2}
\label{tab3}\begin{tabular*}{\textwidth}{@{\extracolsep{4in minus 4in}}lcd{2.14}d{2.14}d{2.14}@{\hspace*{-1.5pt}}}
\hline
& & \multicolumn{1}{c}{\textbf{Complete-case analysis}} & \multicolumn{1}{c}{\textbf{Available-cases analysis}}
& \multicolumn{1}{c@{}}{\textbf{Proposed
method}}\\ \hline
 {Patients} & Intercept & 2.53 \ (-1.71, 6.77)& 0.99
\ (-2.55, 4.52)& 5.10 \ (3.31, 6.59) \\
& Patient CESD & 0.43 \ (0.29, 0.58) & 0.56 \ (0.44, 0.68)&0.87 \ (0.80,
0.93)\\
& Spouse CESD & 0.07 \ (-0.06, 0.20) & 0.06 \ (-0.06, 0.17) &0.14 \ (0.09,
0.19)\\
& Patient MPI &0.94 \ (0.22, 1.67) & 0.82 \ (0.21, 1.43) &1.24 \ (0.83,
1.64)\\
& Spouse MPI & 1.06 \ (0.29, 1.82) &0.90 \ (0.31, 1.48) &0.62 \ (0.40, 0.84)\\
& Cancer stage & 0.39 \ (-0.81, 1.60) &0.59 \ (-0.43, 1.60) &0.10 \ (-0.47,
0.66) \\
[4pt]
 Spouses  &Intercept & 3.68 \ (-0.55, 7.92)& 2.00 \ (-1.63,
5.64) &8.16 \ (4.26, 11.9) \\
& Patient CESD & -0.05 \ (-0.19, 0.09) & 0.01 \ (-0.11, 0.13) & 0.68 \ (0.63,
0.74) \\
& Spouse CESD & 0.77 \ (0.64, 0.90) &0.78 \ ( 0.66, 0.89) & 0.76 \ (0.71,
0.81)\\
&Patient MPI & 0.43 \ (-0.29, 1.15) &0.27 \ (-0.27, 0.81) &0.53 \ (0.33,
0.73)\\
& Spouse MPI & 0.55 \ (-0.22, 1.31) &0.58 \ (-0.04, 1.20)&0.36 \ (-0.64,
1.15)\\
&Cancer stage & -0.42 \ (-1.63, 0.79)&-0.21 \ (-1.23, 0.80)&-0.50 \ (-0.92,
0.09)\\
\hline
\end{tabular*}
\end{table}

As shown in Table~\ref{tab3}, the proposed method suggests significant
``partner'' effects for the patients. Specifically, the patient's
depression increases with her spouse's MPI [$\mathrm{estimate} = 0.62$ and 95\% $\mathrm{CI}
= (0.40, 0.84)$] and previous CESD [$\mathrm{estimate} = 0.14$ and 95\% $\mathrm{CI} = (0.09, 0.19)$].
In addition, there are also significant ``actor'' effects for the
patients, that is, the patient's depression is positively correlated
with her own MPI and previous CESD scores.
For the spouses, we observed similar significant ``partner'' effects:
the spouse's depression increases with the patient's MPI and previous
CESD scores. However, the ``actor'' effects for the spouses are
different from those for the patients. The spouse's depression
correlates with his previous CESD scores but not the MPI level, whereas
the patient's depression is related to both variables. Based on these
results, we can see that the patients and spouses are highly
interdependent and influence each other's depression status. Therefore,
when designing a prevention program to reduce depression in patients,
we may achieve better outcomes by targeting both patients and spouses
simultaneously.

As for the dropout process, the results in Table~\ref{tab4} suggest that the
missing data for the patients are nonignorable because the  probability
of dropout is significantly associated with the patient's current
(missing) CESD score. In contrast, the missing data for the spouse
appears to be ignorable, as the probability of dropout does not depend
on the spouse's current (missing) CESD score.
For the variance components, the estimates of residuals variances for
patients and spouses are $\hat{\sigma}_1^2=5.02$ [95\% $\mathrm{CI} = (2.98,
7.01)$] and $\hat{\sigma}_2^2=6.12$ [95\% $\mathrm{CI} = (4.03, 7.95)$],
respectively. The estimates of the variances for the random effects}
$b_i$ and $c_i$ are $\hat{\tau}_b^2=9.95$ [95\% $\mathrm{CI} = (7.96, 11.92)$]
and $\hat{\tau}_c^2=7.97$ [95\% $\mathrm{CI} = (5.99, 9.89)$], respectively,
suggesting substantial variations across dyads.

%t4 ###
%
%t3 #&#
\begin{table}
\tabcolsep=0pt
\caption{Parameter estimates and 95\% credible intervals (shown in
parentheses) of the dropout model for the breast cancer data}\label{tbProbitModel2}
{\fontsize{8.5pt}{10.5pt}\selectfont{\label{tab4}\begin{tabular*}{\textwidth}{@{\extracolsep{\fill}}ld{3.13}d{2.12}d{2.11}d{2.11}d{2.11}@{\hspace*{-1.5pt}}}
\hline
 & \multicolumn{1}{c}{\textbf{Intercept}} & \multicolumn{1}{c}{\textbf{Current CESD}}
 & \multicolumn{1}{c}{\textbf{Previous CESD}}& \multicolumn{1}{c}{\textbf{MPI}} & \multicolumn{1}{c@{}}{\textbf{Cancer stage}} \\
\hline
   Patients & -0.8 \ (-8.3, 6.2) & -1.6 \ (-4.2, -0.3) & 0.6 \ (-0.3, 1.6) & 0.8 \ (-1.6, 3.8) & -0.4 \ (-0.9, 2.4) \\
Spouses & -15.6 \ (-25.6, -4.1) & 0.8 \ (-0.2, 1.6) & -0.7 \ (-1.6, 0.5) &
-0.2 \ (-2.1, 1.4)& 2.9 \ (-1.7, 6.4) \\
\hline
\end{tabular*}}}
\end{table}

Compared to the proposed approach, both the complete-case and
available-case analyses fail to detect some ``partner'' effects. For
example, for spouses, the complete-case and available-case analyses
assert that the spouse's CESD {is correlated with} his own previous
CESD scores only, whereas the proposed method suggested that the
spouse's CESD is {related} not only to his own CESD but also to the
patient's CESD and MPI level. In addition, for patients, the
``partner''
effect of the spouse's CESD is not significant under the complete-case
and available-case analyses, but is significant under the proposed
approach. These results suggest that ignoring the nonignorable dropouts
could lead to a failure to detect important covariate effects.

\begin{figure}

\includegraphics{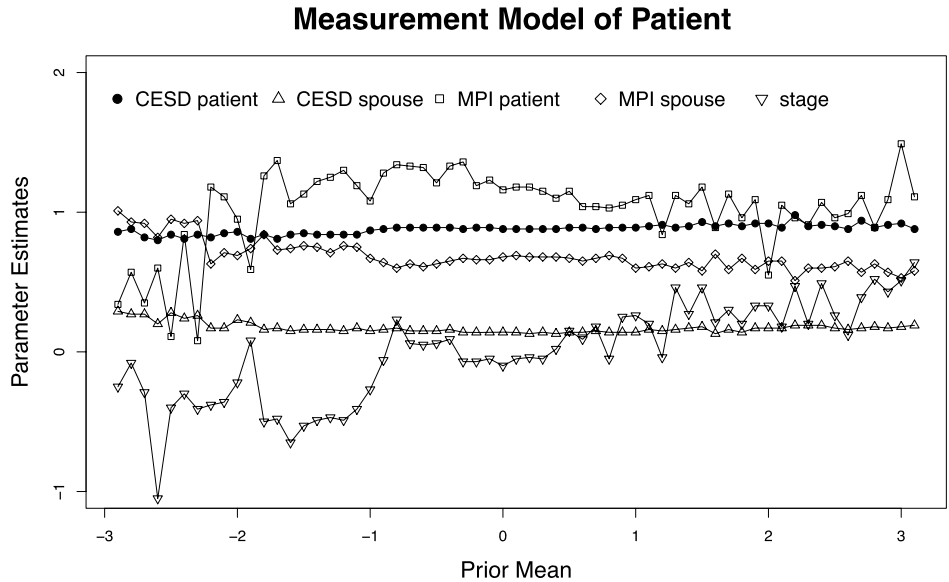}

\caption{Sensitivity analysis of the proposed nonignorable model for
the breast cancer data. The figure shows the parameter estimates of the
patients' measurement model under informative normal priors for $\phi
_1$ and $\phi_2$ with a mean varying from $-3$ to 3 and a fixed variance
of 0.01.}\label{figSensitivityGibbs1}
\label{fig1}\end{figure}

\begin{figure}

\includegraphics{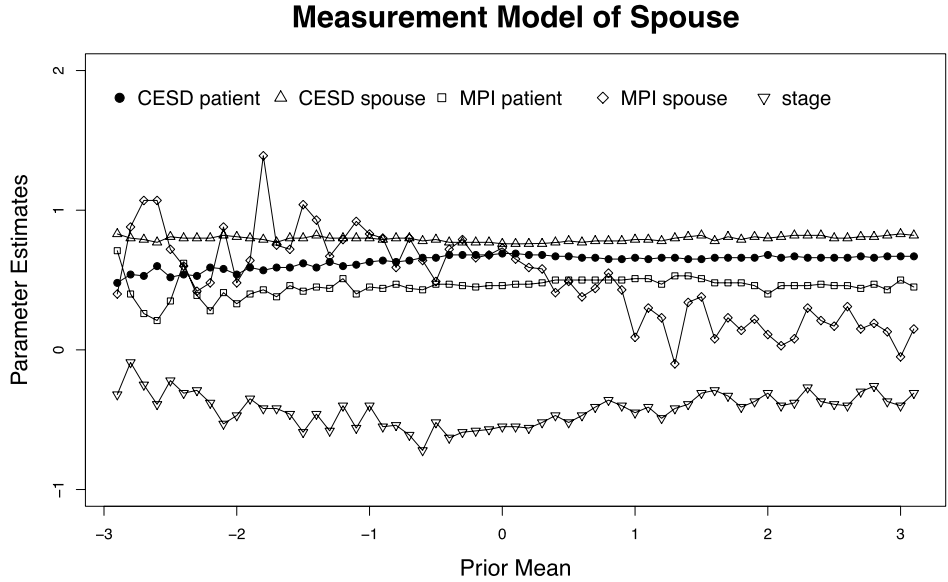}

\caption{Sensitivity analysis of the proposed nonignorable model for
the breast cancer data. The figure shows the parameter estimates of the
spouses' measurement model under informative normal priors for $\phi
_1$ and $\phi_2$ with a mean varying from $-3$ to 3 and a fixed variance
of 0.01.}\label{figSensitivityGibbs2}
\label{fig2}\end{figure}

Nonidentifiability is a common problem when modeling nonignorable
missing data. In our approach, the observed data contain very limited
information on the parameters that link the missing outcome with the
dropout process, that is, $\phi_1$ and $\phi_2$ in the dropout model.
The identification of these parameters is heavily driven by the
untestable model assumptions [\citet{VerMol00}; \citet{LitRub02}]. In this case, a
sensible strategy is to perform a sensitivity analysis to examine how
the inference changes with respect to the values of $\phi_{1}$ and
$\phi_{2}$ [Daniels and Hogan (\citeyear{DanHog00}, \citeyear{DanHog08});
\citet{Rotetal01}]. We conducted a Bayesian sensitivity analysis by
assuming informative normal prior distributions for $\phi_1$ and $\phi
_2$ with a small variance of 0.01 and the mean fixed, successively, at
various values. Figures \ref{figSensitivityGibbs1} and \ref
{figSensitivityGibbs2} show the parameter estimates of the measurement
models when the prior means of $\phi_1$ and $\phi_2$ vary from $-3$ to 3.
In general, the estimates were quite stable, except that the estimate
of cancer stage in the measurement model of patient (Figure \ref
{figSensitivityGibbs1}) and the estimate of spouse's MPI in the
measurement model of spouse (Figure \ref{figSensitivityGibbs2})
demonstrated some variations.

%f1 ###
%
%f1 #&#

%f2 ###
%
%f2 #&#

We conducted another sensitivity analysis on the prior distributions of~$\sigma_1^2$, $\sigma_2^2$, $\tau_b^2$ and $\tau_c^2$. We
considered various inverse gamma priors, $\mathit{IG}(a, b)$, by setting $a = b
= 0.01, 1 $ and 5. As shown in Table \ref{tablepriorvar}, the
estimates of the measurement model parameters were stable under
different prior distributions, suggesting the proposed method is not
sensitive to the priors of these parameters.

%s7 ###
%s7 #&#
\section{Conclusion}\label{sec7}\label{seccon}
We have developed a selection-model-based approach to analyze
longitudinal dyadic data with nonignorable dropouts. We model the
longitudinal outcome process using a transition model and account for
the correlation within dyads using random effects. In the model, we
allow a~subject's outcome to depend on not only his/her own
characteristics but also the characteristics of the other member in the
dyad. As a~result, the parameters of the proposed model have appealing
interpretations as ``actor'' and ``partner'' effects, which greatly
facilitates the understanding of interdependence within a relationship
and the design of more efficient prevention programs. To account for
the nonignorable dropout, we adopt a~discrete time survival model to
link the dropout process with the longitudinal measurement process. We
used the data augment method to address the complex missing data
problem caused by dropout and interdependence within dyads. The
simulation study shows that the proposed method yields consistent
estimates with correct coverage rates. We apply our methodology to the
longitudinal dyadic data collected from a breast cancer study. Our
method identifies more ``partner'' effects than the methods that ignore
the missing data, thereby providing extra insights into the
interdependence of the dyads. For example, the methods that ignore the
missing data suggest that the spouse's CESD related only to his own
previous CESD scores, whereas the proposed method suggested that the
spouse's CESD related not only to his own CESD but also to the
patient's CESD and MPI level. This extra information can be useful for
the design of more efficient depression prevention programs for breast
cancer patients.

%t5 ###
%
%t4 #&#
\begin{table}
\tabcolsep=0pt
\caption{Parameter estimates and 95\% credible intervals (show in
parentheses) for the patient's and spouse's measurement models by
fixing $a$ and $b$ at 0.01, 1 and 5 for the inverse gamma prior $\mathit{IG} (a,
b)$ on $\sigma_1^2$, $\sigma_2^2$, $\tau_b^2$ and $\tau_c^2$}\label{tablepriorvar}
\label{tab5}\begin{tabular*}{\textwidth}{@{\extracolsep{\fill}}lcd{2.14}d{2.14}d{2.14}@{\hspace*{-1.5pt}}}
\hline
& & \multicolumn{1}{c}{$\boldsymbol{a=b=0.01}$} & \multicolumn{1}{c}{$\boldsymbol{a=b=1}$} & \multicolumn{1}{c@{}}{$\boldsymbol{a=b=5}$} \\
\hline
 {Patients} & Intercept &4.72 \ ( 3.32, 6.11) &5.00 \ (
3.48, 6.47)&5.02 \ ( 3.57, 6.48) \\
&Patient CESD &0.87 \ ( 0.81, 0.93) &0.86 \ ( 0.80, 0.92) &0.88 \ ( 0.83,
0.94) \\
& Spouse CESD &0.14 \ ( 0.09, 0.19) &0.14 \ ( 0.08, 0.19) &0.13 \ ( 0.08,
0.18) \\
&Patient MPI &1.27 \ ( 0.84, 1.71) &1.12 \ ( 0.67, 1.60) &1.20 \ ( 0.85,
1.57)\\
&Spouse MPI &0.71 \ ( 0.49, 0.91) &0.68 \ ( 0.46, 0.87) &0.61 \ ( 0.39, 0.82)
\\
&Cancer stage & -0.03 \ (-0.50, 0.50)&0.18 \ (-0.31, 0.65)&-0.08 \ (-0.57,
0.40) \\[4pt]
 {Spouses}& Intercept & 6.40 \ ( 4.39, 8.41) &7.56 \ ( 5.35,
9.93) &7.52 \ ( 5.43, 9.55) \\
&Patient CESD &0.67 \ ( 0.62, 0.73) &0.67 \ ( 0.62, 0.72) &0.69 \ ( 0.64,
0.73) \\
& Spouse CESD &0.76 \ ( 0.71, 0.80) &0.75 \ ( 0.71, 0.81) &0.75 \ ( 0.71,
0.80) \\
&Patient MPI &0.51 \ ( 0.32, 0.71) &0.54 \ ( 0.35, 0.73) &0.53 \ ( 0.34,
0.72)\\
&Spouse MPI &0.79 \ (-0.05, 1.46) &0.54 \ (-0.03, 1.06) &0.45 \ (-0.23, 1.09)
\\
&Cancer stage & -0.41 \ (-0.86, 0.02) &-0.38 \ (-0.81, 0.03) &-0.48 \ (-0.87,
0.08)\\ \hline
\end{tabular*}
\end{table}

In the proposed dropout model (\ref{dropmodel}), we assume that
time-dependent covariates $\bs{X}_{kij}$ and ${Y}_{kij}$, $k=1, 2$,
have captured all important time-dependent factors that influence
dropout. However, this assumption may not be always true. A more
flexible approach is to include in the model a time-dependent random
effect $c_{ij}$ to represent all unmeasured time-variant factors that
influence dropout. We can further put a hierarchical structure on
$c_{ij}$ to shrink it toward a dyad-level time-invariant random effect
$c_i$ to account for the effects of unmeasured time-invariance factors
on dropout. In addition, in (\ref{dropmodel}), in order to allow
members in a dyad to drop out at different times, we specify separate
dropout models for each dyadic member, linked by a common random
effect. Although the common random effect makes the members in a dyad
more likely to drop out at the same time, it may not be the most
effective modeling approach when dropout mostly occurs at the dyad
level. In this case, a more effective approach is that, in addition to
the dyad-level random effect, we further put hierarchical structure on
the coefficients of common covariates (in the two dropout models) to
shrink toward a common value to reflect that dropout is almost always
at the dyad level.

\section*{Acknowledgments}
We would like to thank the referees, Associate Editor and Editor
(Professor Susan Paddock) for very helpful comments
that substantially improved this paper.

%suskaldyti doi

% imsref loaded by smiklovaite, 2012-01-03 08:34:44
%

\printaddresses

\end{document}